\documentclass[runningheads]{svmult}

\usepackage{makeidx}   % allows index generation
\usepackage{graphicx}  % standard LaTeX graphics tool
\usepackage{subeqnar}  % subnumbers individual equations
\usepackage{multicol}  % used for the two-column index
\usepackage{physmubb}  % centered layout of captions etc.
\makeindex             % used for the subject index
\usepackage{psfig}
\usepackage{epsfig}
\usepackage{amssymb,wasysym,bbm,feynmf,rotating,subfigure,here,citesort}
\newcommand{\alphaS}{\alpha_s}
\newcommand{\alphaEM}{\alpha}

\newcommand{\pert}{P}
\newcommand{\nprt}{N\!P}

\newcommand{\pbar}{{\bar{p}}}

\newcommand{\be}{\begin{equation}}
\newcommand{\ee}{\end{equation}}
\newcommand{\bea}{\begin{eqnarray}}
\newcommand{\eea}{\end{eqnarray}}
\newcommand{\benn}{\begin{displaymath}}
\newcommand{\eenn}{\end{displaymath}}
\newcommand{\beann}{\begin{eqnarray*}}
\newcommand{\eeann}{\end{eqnarray*}}
\newcommand{\barray}{\begin{array}}
\newcommand{\earray}{\end{array}}
\newcommand{\inv}{\frac{1}}
\newcommand{\gtsim}{\;\lower-0.45ex\hbox{$>$}\kern-0.77em\lower0.55ex\hbox{$\sim$}\;}
\newcommand{\ltsim}{\;\lower-0.45ex\hbox{$<$}\kern-0.77em\lower0.55ex\hbox{$\sim$}\;}

\newcommand{\GeV}{\mbox{GeV}}
\newcommand{\TeV}{\mbox{TeV}}
\newcommand{\G}{{\cal G}}       % Gluon Field Strength
\newcommand{\Pc}{{\cal P}}      % Path Ordering
\newcommand{\Tr}{\mbox{Tr}}    % Trace
\newcommand{\im}{\mbox{Im}}    % Imaginary Part
\newcommand{\befig}{\begin{figure}}
\newcommand{\efig}{\end{figure}}
\newcommand{\betab}{\begin{table}}
\newcommand{\etab}{\end{table}}
\begin{document}
%
%
%
%%%%%%%%%%% HD Titlepage %%%%%%%%%%%

\begin{titlepage}
\begin{flushright}
\begin{tabular}{l}
HD-THEP-02-41\\
hep-ph/0212070\\
% December 2002
\end{tabular}
\end{flushright}

\vspace*{1.3truecm}

\begin{center}
\boldmath
{\Large \bf Saturation Effects in Hadronic Cross Sections$^*$}\\ 
\vspace*{0.2truecm}
%our Knowledge of the Standard Model}
\unboldmath

\vspace*{1.6cm}

\smallskip
\begin{center}
{\sc 
{\large Arif~I.~Shoshi
%\footnote{\tt shoshi@tphys.uni-heidelberg.de} 
and Frank~D.~Steffen
%\footnote{\tt Frank.D.Steffen@thphys.uni-heidelberg.de}
}}\\
\vspace*{2mm}
{\sl Institut f\"ur Theoretische Physik, Universit\"at Heidelberg\\
Philosophenweg 16 \& $\!$19, D-69120 Heidelberg, Germany}
%E-mail: {\tt Frank.D.Steffen@thphys.uni-heidelberg.de}}
\end{center}

\vspace{2.0truecm}

{\large\bf Abstract\\[10pt]} \parbox[t]{\textwidth}{ 
  We compute total and differential elastic cross sections of
  high-energy hadronic collisions in the loop-loop correlation model
  that provides a unified description of hadron-hadron, photon-hadron,
  and photon-photon reactions. The impact parameter profiles of $pp$
  and $\gamma^*p$ collisions are calculated. For ultra-high energies
  the hadron opacity saturates at the black disc limit which tames the
  growth of the hadronic cross sections in agreement with the
  Froissart bound.  We compute the impact parameter dependent gluon
  distribution of the proton $xG(x,Q^2,|\vec{b}_{\perp}|)$ and find
  gluon saturation at small Bjorken $x$. These saturation effects
  manifest S-matrix unitarity in hadronic collisions and should be
  observable in future cosmic ray and accelerator experiments at
  ultra-high energies. The c.m.\ energies and Bjorken $x$ at which
  saturation sets in are determined and LHC and THERA predictions are
  given.
}

\vspace{1.5cm}
 
$^*${\sl Talk presented by F.~D.~Steffen at the 14th Topical
  Conference on Hadron Collider Physics, Karlsruhe, Germany, September
  29 -- October 4, 2002}
%\\
%To appear in the Proceedings}
\end{center}

\end{titlepage}
 
\thispagestyle{empty}
\vbox{}
\newpage
 
\setcounter{page}{1}
 
%%% end HD title page %%%%%%%%%%%%%
%
%
%
\title*{Saturation Effects in Hadronic Cross Sections}
\toctitle{Saturation Effects in Hadronic Cross Sections}
\titlerunning{Saturation Effects in Hadronic Cross Sections}
% allows abbreviation of title, if the full title is too long
% to fit in the running head
%
\author{Arif~I.~Shoshi and \underline{Frank~D.~Steffen}\\
Institut f\"ur Theoretische Physik, Universit\"at Heidelberg\\
Philosophenweg 16 \& $\!$19, D-69120 Heidelberg, Germany}
\authorrunning{Arif~I.~Shoshi and Frank~D.~Steffen}
% if there are more than two authors,
% please abbreviate author list for running head
%

\maketitle              % typesets the title of the contribution

% ___ Introduction _____________________________________________________________
\section{Introduction}
\label{Sec_Introduction}
% ______________________________________________________________________________

The steep rise of the gluon distribution $xG(x,Q^2)$ and structure
function $F_2(x,Q^2)$ of the proton towards small Bjorken $x=Q^2/s$ is
one of the most exciting observations at the HERA
experiments~\cite{Adloff:1997mf+X}. The experimental results show a
rise of the total $\gamma^* p$ cross section,
$\sigma_{\gamma^*p}^{tot}(s,Q^2)$, with increasing c.m.\ energy
$\sqrt{s}$ which becomes stronger with increasing photon virtuality
$Q^2$. In hadronic interactions, the rise of the total cross sections
is limited: The Froissart bound, derived from very general principles
such as unitarity and analyticity of the $S$-matrix, allows at most a
logarithmic energy dependence of the cross sections at asymptotic
energies~\cite{Froissart:1961ux}.  Analogously, the rise of
$\sigma_{\gamma^* p}^{tot}(s,Q^2)$ is expected to slow down. The
microscopic picture behind this slow-down is the concept of gluon
saturation: Since the gluon density in the proton becomes large at
high energies $\sqrt{s}$ (small $x$), gluon fusion processes are
expected to tame the growth of $\sigma_{\gamma^* p}^{tot}(s,Q^2)$, and
it is a key issue to determine the energy at which these processes
become significant.

In this talk we give predictions for saturation effects in hadronic
cross sections using the loop-loop correlation model (LLCM) that
provides a unified description of hadron-hadron, photon-hadron, and
photon-photon reactions~\cite{Shoshi:2002in}. The saturation effects
are in agreement with $S$-matrix unitarity constraints and result from
multiple gluonic interactions between the scattered particles. We show
how these manifestations of $S$-matrix unitarity can in principle be
observed in future cosmic ray and accelerator experiments at
ultra-high energies. The c.m.\ energies and Bjorken $x$ at which
saturation sets in are determined and LHC and THERA predictions are
given. The presented results are extracted from~\cite{Shoshi:2002in}
where more details can be found.

\medskip

% ______________________________________________________________________________
\section{The Loop-Loop Correlation Model}
\label{Sec_The_Model}
% ______________________________________________________________________________

The loop-loop correlation model (LLCM)~\cite{Shoshi:2002in} is based
on the functional integral approach to high-energy
collisions~\cite{Nachtmann:1991ua,Dosch:1994ym,Nachtmann:ed.kt,Dosch:RioLecture}.
Accordingly, the $T$-matrix element for elastic proton-proton ($pp$)
scattering at c.m.\ energy squared $s$ and transverse momentum
transfer $\vec{q}_{\perp}$ ($t = -\vec{q}_{\!\perp}^{2}$) reads
\bea
        T_{pp}(s,t) 
        & = & \,\,
        2\I s \!\!\int \!\!d^2b_{\!\perp}\,
        e^{\I {\vec q}_{\!\perp} {\vec b}_{\!\perp}}\,
        J_{pp}(s,|\vec{b}_{\!\perp}|)  
\label{Eq_T_pp_matrix_element} \\        
        J_{pp}(s,|\vec{b}_{\!\perp}|)  
        & = & 
        \int \!\!dz_1 d^2r_1\!\! \int \!\!dz_2 d^2r_2      
        |\psi_p(z_1,\vec{r}_1)|^2 |\psi_p(z_2,\vec{r}_2)|^2
\nonumber\\ &&
        \times 
        \left[1-S_{DD}(s,{\vec b}_{\!\perp},z_1,{\vec r}_1,z_2,{\vec r}_2)\right]
\label{Eq_model_pp_profile_function}
\eea
where the correlation of two light-like Wegner-Wilson loops
\be
        S_{DD}(s,{\vec b}_{\!\perp},...)
        = \Big\langle\!W[C_1] W[C_2]\!\Big\rangle_G
        \,
        \mbox{with}
        \,\,
        W[C_{i}] = 
        \inv{3} \Tr\,\Pc
        \exp\!\Big[\!-\I g\!\oint_{C_{i}}\!\!\!\!dz^{\mu}\G_{\mu}(z)\Big]    
\label{Eq_S_DD_def_W[C]_def}
\ee
describes the elastic scattering of two light-like color dipoles
($DD$) with transverse size and orientation $\vec{r}_{\!i}$ and
longitudinal quark momentum fraction $z_i$ at impact parameter
$\vec{b}_{\!\perp}$, i.e.\ the loops $C_i$ represent the trajectories
of the scattering color dipoles. For elastic $pp$ scattering, the
color dipoles are given in a simplified picture by a quark and diquark
in each proton with ${\vec r}_i$ and $z_i$ distributions described by
the simple phenomenological Gaussian wave function
$|\psi_p(z_i,\vec{r}_i)|^2$. For reactions involving (virtual)
photons, the quark and antiquark in the photon form a color dipole
whose ${\vec r}_i$ and $z_i$ distribution is described by the
perturbatively derived photon wave function
$|\psi_{\gamma^*_{T,L}}(z_i,\vec{r}_i,Q^2)|^2$.  To account for the
non-perturbative region of low $Q^2$ in the photon wave function,
quark masses $m_f(Q^2)$ are used that interpolate between current
quarks at large $Q^2$ and constituent quarks at small
$Q^2$~\cite{Dosch:1998nw}.

In contrast to the wave functions, the loop-loop correlation function
$S_{DD}$ is universal for $pp$, $\gamma^* p$, and $\gamma\gamma$
reactions~\cite{Shoshi:2002in}. We compute $S_{DD}$ in the
Berger-Nachtmann approach~\cite{Berger:1999gu}, in which $S$-matrix
unitarity is respected, and describe the QCD interactions between the
color dipoles by combining the non-perturbative stochastic vacuum
model~\cite{Dosch:1987sk+X} with perturbative gluon exchange.  This
combination allows us to describe long and short distance correlations
in agreement with numerical lattice computations~\cite{Bali:1998aj+X}
and leads to the static quark-antiquark potential with color Coulomb
behavior for small source separations and confining linear rise for
large source separations~\cite{Shoshi:2002rd}. Two components are
obtained of which the perturbative ($\pert$) component,
$(\chi^{\pert})^2$, describes two-gluon exchange and the
non-perturbative ($\nprt$) component, $(\chi^{\nprt})^2$, the
corresponding non-perturbative two-point
interaction~\cite{Shoshi:2002fq}. Ascribing a weak ($\epsilon^{\nprt}
= 0.125$) and strong ($\epsilon^{\pert} = 0.73$) powerlike energy
dependence to the non-perturbative and perturbative component,
respectively,
\be
        \chi^{\nprt}\!(s)^{2} = \left(\chi^{\nprt}\right)^{\!2} \!
        \left(\!\frac{s}{s_0}
        \frac{\vec{r}_1^{\,2}\,\vec{r}_2^{\,2}}{R_0^4}\right)^{\!\!\epsilon^{\nprt}}
        \,\,\mbox{and}\,\,\,\,\,\,\,
        \chi^{\pert}\!(s)^{2} = \left(\chi^{\pert}\right)^{\!2} \!
        \left(\!\frac{s}{s_0} 
        \frac{\vec{r}_1^{\,2}\,\vec{r}_2^{\,2}}{R_0^4}\right)^{\!\!\epsilon^{\pert}}
        \ ,
\label{Eq_energy_dependence}
\ee
our final result for $S_{DD}$ reads
\be
        S_{DD}
        = \frac{2}{3} 
        \cos\!\left[\frac{1}{3}\chi^{\nprt}(s)\right]
        \cos\!\left[\frac{1}{3}\chi^{\pert}(s)\right]         
        + \frac{1}{3}
        \cos\!\left[\frac{2}{3}\chi^{\nprt}(s)\right]
        \cos\!\left[\frac{2}{3}\chi^{\pert}(s)\right]
        \ .
\label{Eq_S_DD_final_result}
\ee
The cosine functions ensure the unitarity condition in impact
parameter space as they average to zero in the integration over the
dipole orientations at very high energies. In fact, the higher order
terms in the expansion of the cosine functions describe multiple
gluonic interactions that are crucial for the saturation effects shown
below: They tame the rise of hadronic cross sections and the gluon
distribution in the proton $xG(x,Q^2)$ at ultra-high
energies~\cite{Shoshi:2002in}.

% ______________________________________________________________________________
\section{Saturation in Proton-Proton Scattering}
\label{Sec_Saturation_Hadronic_Reactions}
% ______________________________________________________________________________

The profile function~(\ref{Eq_model_pp_profile_function}) is a measure
for the blackness or opacity of the interacting protons and gives an
intuitive geometrical picture for the energy dependence of $pp$
reactions: As shown in Fig.~\ref{Fig_J_pp(b,s)}, the protons become
blacker and larger with increasing c.m.\ energy $\sqrt{s}$. At
ultra-high energies, $\sqrt{s} \gtsim 10^6\,\GeV$, the opacity
saturates at the black disc limit first for zero impact parameter
while the transverse expansion of the protons continues. For purely
imaginary elastic amplitudes, expected at high energies, the black
disc limit is a strict unitarity bound that limits the height of the
profile function at $J_{pp}^{\mbox{\scriptsize max}}\!=\!1$ for proton
wave functions normalized to one
%, $\int \!\!dz_i d^2r_i|\psi_p(z_i,\vec{r}_i)|^2 = 1$
. Thus, the saturation of the profile function is an explicit
manifestation of $S$-matrix unitarity.
\begin{figure}
\begin{center}
\includegraphics[width=.6\textwidth]{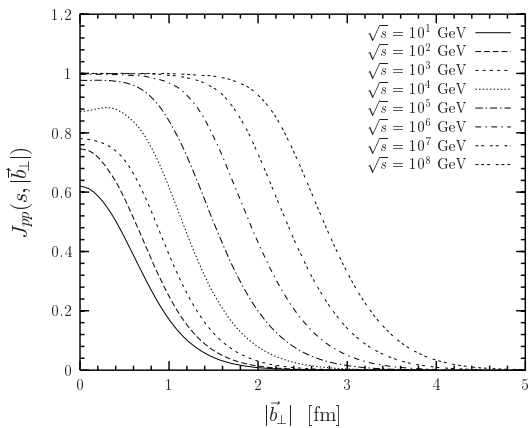}
\end{center}
\caption[]{
  The profile function for proton-proton scattering
  $J_{pp}(s,|\vec{b}_{\!\perp}|)$ as a function of the impact
  parameter $|\vec{b}_{\!\perp}|$ for c.m.\ energies from $\sqrt{s} =
  10\,\GeV$ to $10^8\,\GeV$}
\label{Fig_J_pp(b,s)}
\end{figure}

\begin{figure}
\begin{center}
\includegraphics[width=.8\textwidth]{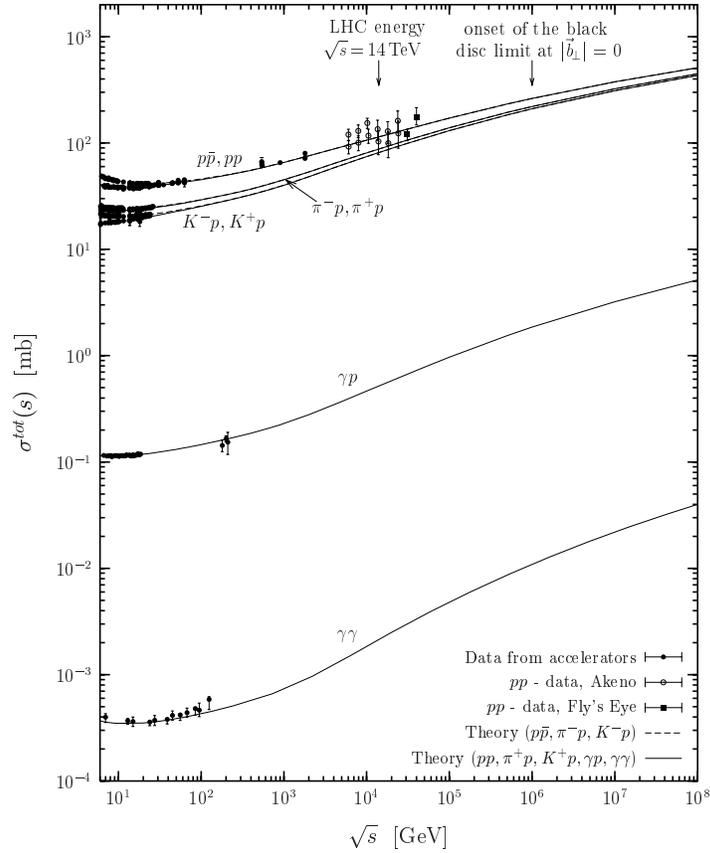}
\end{center}
\caption[]{
  The total cross section $\sigma^{tot}$ as a function of the c.m.\ 
  energy $\sqrt{s}$ for $pp$, $p\pbar$, $\pi^{\pm}p$, $K^{\pm}p$,
  $\gamma p$ and $\gamma \gamma$ scattering.  The solid lines
  represent the LLCM results for $pp$, $\pi^+p$, $K^+p$, $\gamma p$
  and $\gamma \gamma$ scattering and the dashed lines the ones for
  $p\pbar$, $\pi^-p$, and $K^-p$ scattering. The $pp$, $p\pbar$,
  $\pi^{\pm}p$, $K^{\pm}p$, $\gamma p$~\cite{Groom:2000in} and $\gamma
  \gamma$ data~\cite{Abbiendi:2000sz+X} taken at accelerators are
  indicated by the closed circles while the closed squares (Fly's eye
  data)~\cite{Baltrusaitis:1984ka+X} and the open circles (Akeno
  data)~\cite{Honda:1993kv+X} indicate cosmic ray data}
\label{Fig_sigma_tot}
\end{figure}

The total $pp$ cross section is obtained directly from the profile
function~(\ref{Eq_model_pp_profile_function})
\be
        \sigma^{tot}_{pp}(s) 
        \,=\, \inv{s}\,\im\,T_{pp}(s, t\!=\!0) 
        \,=\, 2 \int \!d^2b_{\!\perp}\,J_{pp}(s,|\vec{b}_{\!\perp}|)
        \ .
\label{Eq_optical_theorem}
\ee
and shows saturation effects in principle observable in experiments.
As can be seen in Fig.~\ref{Fig_sigma_tot}, the saturation of
$J_{pp}(s,|\vec{b}_{\!\perp}|)$ tames the growth of
$\sigma^{tot}_{pp}(s)$: There is a transition from a power-like to an
$\ln^2$-increase of $\sigma^{tot}_{pp}(s)$, which respects the
Froissart bound~\cite{Froissart:1961ux}, at about $\sqrt{s} \approx
10^3\,\TeV$. Thus -- according to our model -- the onset of the black
disc limit in $pp$ collisions is about two orders of magnitude beyond
LHC energy $\sqrt{s} = 14\,\TeV$ and clearly out of reach for
accelerator experiments in the near future. Here cosmic ray
experiments might help that have access to energies of up to about
$10^{8}\,\TeV$.

\begin{figure}
\begin{center}
\includegraphics[width=.75\textwidth]{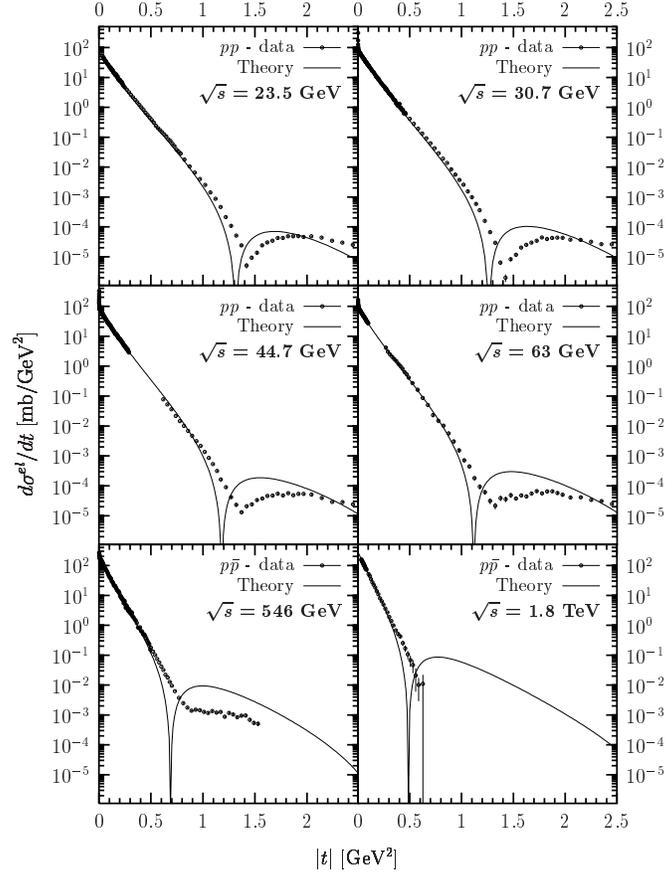}
\end{center}
\caption[]{
  The differential elastic cross section for $pp$ and $p\pbar$
  scattering as a function of the squared momentum transfer $|t|$. The
  LLCM results ({\it solid line}\,) are compared for $\sqrt{s} =
  23.5,\,30.7,\,44.7\mbox{ and }63\,\GeV$ to the CERN ISR $pp$
  data~\cite{Amaldi:1980kd}, for $\sqrt{s} = 546\,\GeV$ to the CERN
  $Sp{\pbar}S$ data~\cite{Bozzo:1984ri}, and for $\sqrt{s} =
  1.8\,\TeV$ to Fermilab Tevatron $p\pbar$
  data~\cite{Amos:1989at,Amos:1990jh} ({\it open circles}\,)}
\label{Fig_dsigma_el_dt_pp}
\end{figure}

The evolution of the profile function towards its saturation at the
black disc limit is already interesting below $\sqrt{s} \approx
10^3\,\TeV$. Here the key quantity is the differential elastic cross
section which is obtained for purely imaginary $T$-matrix elements by
Fourier transforming the profile
function~(\ref{Eq_model_pp_profile_function})
\be
        \frac{d\sigma^{el}_{pp}}{dt}(s,t) 
        = \inv{16 \pi s^2}|T_{pp}(s,t)|^2
        = \inv{4\pi} \left[ 
        \int \!\!d^2b_{\!\perp}\,
        e^{i {\vec q}_{\!\perp} {\vec b}_{\!\perp}}\,
        J_{pp}(s,|\vec{b}_{\!\perp}|)
        \right ]^2
        \ .
\label{Eq_dsigma_el_dt_model}
\ee
Thus, the agreement of our model results with the data up to
$\sqrt{s}=1.8\,\TeV$ shown in Fig.~\ref{Fig_dsigma_el_dt_pp} is an
important verification of the profiles shown in
Fig.~\ref{Fig_J_pp(b,s)} up to $\sqrt{s} = 10^3\,\GeV$.
The deviations from the data in the dip region are not surprising
since we work with a purely imaginary $T$-matrix
element~\cite{Shoshi:2002in}. A real part is expected to be important
in the dip region which is negligible in comparison to the imaginary
part in the small $|t|$ region. Moreover, our model describes the
pomeron ($C=+1$) contribution but not the odderon ($C=-1$)
contribution important for the difference between $pp$ and $p\pbar$
reactions.

\begin{figure}
\begin{center}
\includegraphics[width=.6\textwidth]{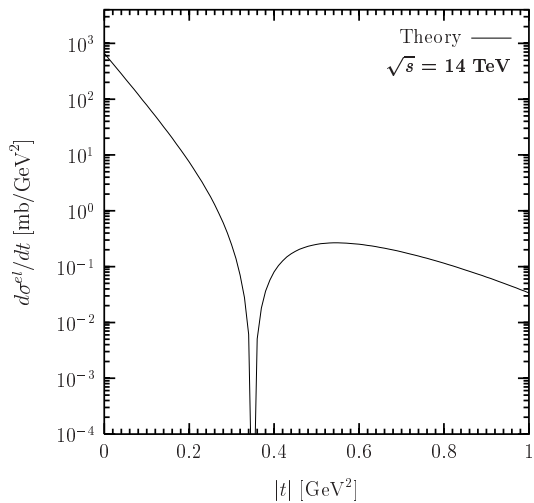}
\end{center}
\caption[]{
  The LLCM prediction of the $pp$ differential elastic cross section
  at LHC ($\sqrt{s} = 14\,\TeV$) as a function of the squared momentum
  transfer $|t|$}
\label{Fig_dsigma_el_dt_pp_LHC}
\end{figure}

It will be very interesting to see the $pp$ differential elastic cross
section measured at LHC and the associated profile function. Our
prediction for $d\sigma^{el}_{\!pp}/dt$ at $\sqrt{s} = 14\,\TeV$ is
shown in Fig.~\ref{Fig_dsigma_el_dt_pp_LHC}.  The associated profile
is close to the dotted line in Fig.~\ref{Fig_J_pp(b,s)} and thus below
the black disc limit.
%

% ______________________________________________________________________________
\section{Gluon Saturation}
\label{Sec_Gluon_Saturation}
% ______________________________________________________________________________

Based on the close relationship between the longitudinal structure
function $F_L(x,Q^2)$ and the gluon distribution of the proton
$xG(x,Q^2)$ at small $x$, the impact parameter dependent gluon
distribution $xG(x,Q^2,|\vec{b}_{\!\perp}|)$ has been related to the
profile function~\cite{Shoshi:2002in}
$J_{\gamma_L^*p}(s=Q^2/x,|\vec{b}_{\!\perp}|,Q^2)$
\be
        xG(x,Q^2,|\vec{b}_{\!\perp}|) 
        \approx
        1.305\,\frac{Q^2}{\pi^2 \alphaS}\,\frac{\pi}{\alphaEM}\,
        J_{\gamma_L^*p}(0.417 x,|\vec{b}_{\!\perp}|,Q^2)
        \ .
\label{Eq_xg(x,Q^2,b)-J_gLp(x,b,Q^2)_relation}
\ee
Consequently, the shape of $xG(x,Q^2,|\vec{b}_{\!\perp}|)$ is
determined by the profile function
$J_{\gamma_L^*p}(s,|\vec{b}_{\!\perp}|,Q^2)$ which is obtained from
(\ref{Eq_model_pp_profile_function}) by replacing
$|\psi_p(z_1,\vec{r}_1)|^2$ with the squared light-cone wave function
for a longitudinally polarized photon
$|\psi_{\gamma_L^*}(z_1,\vec{r}_1,Q^2)|^2$ \cite{Shoshi:2002ri}. Thus,
the blackness described by the profile function is a measure for the
gluon distribution and the black disc limit corresponds to the maximum
of the gluon distribution that can be reached at a given impact
parameter. In accordance with the behavior of the profile function
$J_{\gamma_L^*p}$, see
Fig.~\ref{Fig_J_gp_(b,s,Q^2)_Fig_xg(x,Q^2,b=0)_vs_x}a, the gluon
distribution $xG(x,Q^2,|\vec{b}_{\!\perp}|)$ decreases with increasing
impact parameter for given values of $x$ and $Q^2$. The gluon density,
consequently, has its maximum in the geometrical center of the proton,
i.e.\ at zero impact parameter, and decreases towards the periphery.
With decreasing $x$ at given $Q^2$, the gluon distribution
$xG(x,Q^2,|\vec{b}_{\!\perp}|)$ increases and extends towards larger
impact parameters just as the profile function $J_{\gamma_L^*p}$ for
increasing $s$.  The saturation of the gluon distribution
$xG(x,Q^2,|\vec{b}_{\!\perp}|)$ sets in first in the center of the
proton, $|\vec{b}_{\!\perp}|=0$, at very small Bjorken $x$.
\begin{figure}
\begin{center}
\includegraphics[width=.48\textwidth]{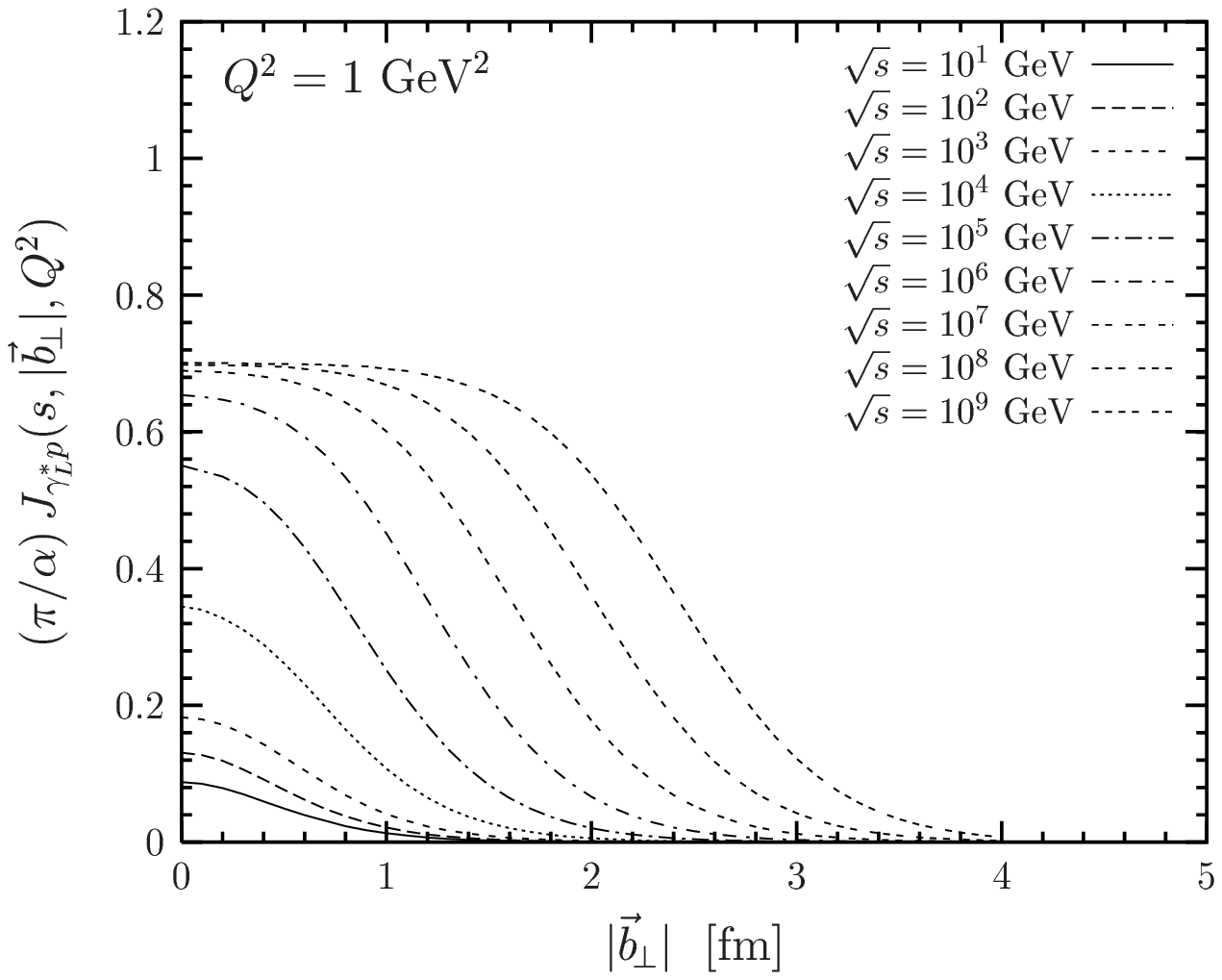}
\hfill
\includegraphics[width=.48\textwidth]{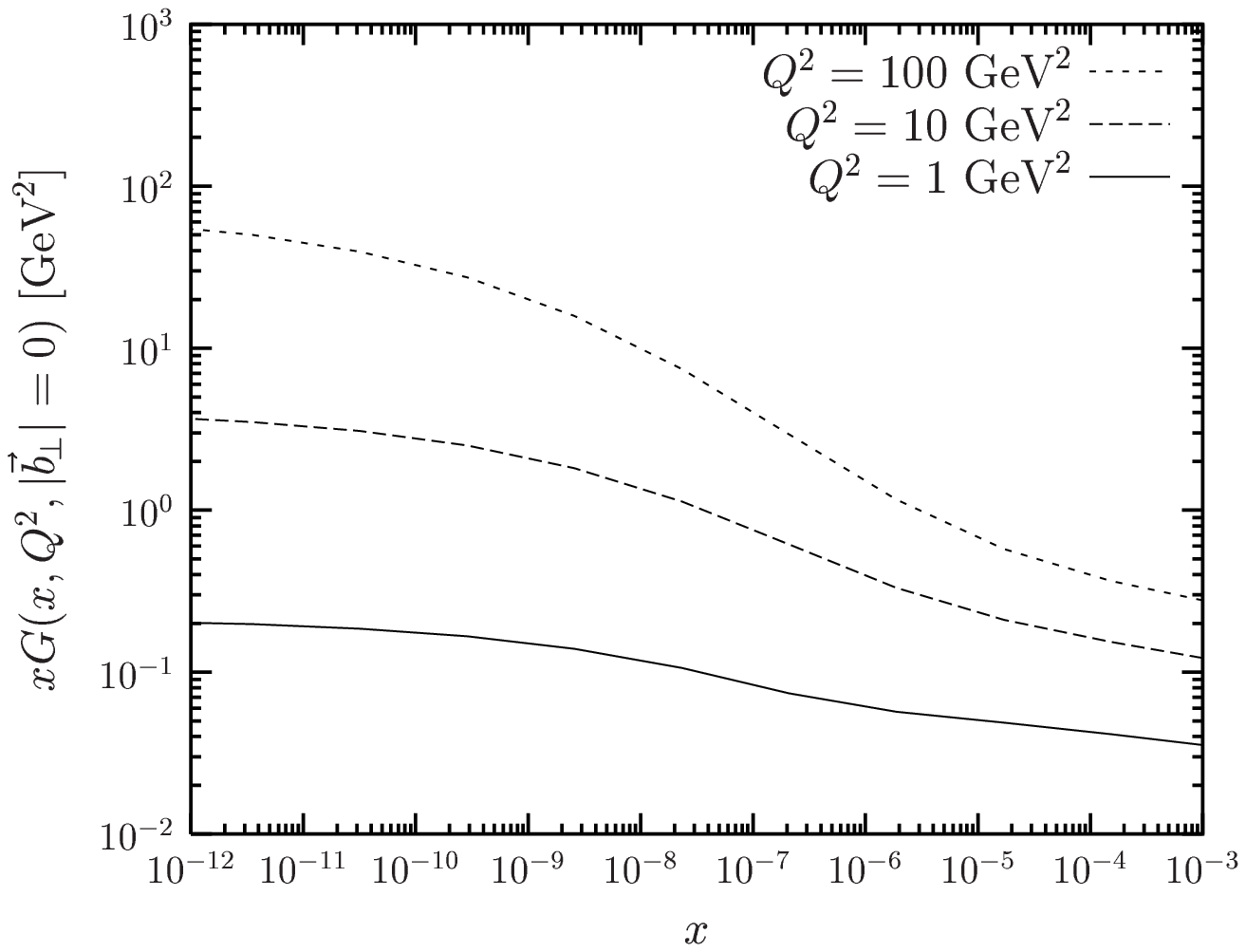}
\end{center}
\caption[]{
  (\textbf{a}) The profile function
  $({\pi}/\alphaEM)J_{\gamma_L^*p}(s,|\vec{b}_{\!\perp}|,Q^2)$ as a
  function of the impact parameter $|\vec{b}_{\!\perp}|$ at photon
  virtuality $Q^2 = 1\,\GeV^2$ for c.m.\ energies from $\sqrt{s} =
  10\,\GeV$ to $10^9\,\GeV$. (\textbf{b}) The gluon distribution of
  the proton at zero impact parameter,
  $xG(x,Q^2,|\vec{b}_{\!\perp}|=0)$, as a function of $x=Q^2/s$ for
  $Q^2 = 1,\,10\mbox{ and }100\,\GeV^2$}
\label{Fig_J_gp_(b,s,Q^2)_Fig_xg(x,Q^2,b=0)_vs_x}
\end{figure}

Using a proton wave function normalized to one, the black
disc limit is given by the normalization of the longitudinal photon
wave function 
\be
        J_{\gamma_L^* p}^{max}(Q^2) 
        = \int \!\!dz d^2r |\psi_{\gamma_L^*}(z,\vec{r},Q^2)|^2
        \ ,
\label{Eq_gp_black_disc_limit}
\ee
which depends on the photon virtuality $Q^2$. This limit
$J_{\gamma^*_L p}^{max}(Q^2)$ induces the upper bound on
$xG(x,Q^2,|\vec{b}_{\!\perp}|)$ and determines the low-$x$ saturation
value
\be
        xG(x,Q^2,|\vec{b}_{\!\perp}|)\ \leq \ 
        xG^{max}(Q^2)
        \approx 
        1.305\,\frac{Q^2}{\pi^2 \alphaS}\,\frac{\pi}{\alphaEM}\,
        J_{\gamma^*_L p}^{max}(Q^2)
        \approx 
        \frac{Q^2}{\pi^2 \alphaS}
        \ ,
\label{Eq_low_x_saturation} 
\ee
which is consistent with complementary
investigations~\cite{Mueller:1986wy+X} and indicates strong color
field strengths $G^a_{\mu \nu} \sim 1/ \sqrt{\alphaS}$ as well. 

In Fig.~\ref{Fig_J_gp_(b,s,Q^2)_Fig_xg(x,Q^2,b=0)_vs_x}b, the
small-$x$ saturation of the gluon distribution at zero impact
parameter $xG(x,Q^2,|\vec{b}_{\!\perp}|=0)$ is illustrated for $Q^2 =
1,\,10\mbox{ and }100\,\GeV^2$ as obtained in the LLCM. The saturation
occurs at very low values of $x \ltsim 10^{-10}$ for $Q^2 \gtsim
1\,\GeV^2$. The photon virtuality $Q^2$ determines the saturation
value (\ref{Eq_low_x_saturation}) and the Bjorken $x$ at which it is
reached: For larger $Q^2$, the small-$x$ saturation value is larger
and is reached at smaller values of $x$. Moreover, the growth of
$xG(x,Q^2,|\vec{b}_{\!\perp}|=0)$ with decreasing $x$ becomes stronger
with increasing $Q^2$. This results from the stronger energy increase
of the perturbative component in the LLCM, $\epsilon^{\pert} = 0.73$,
that becomes more important with decreasing dipole size. In
conclusion, our approach predicts the onset of the
$xG(x,Q^2,|\vec{b}_{\!\perp}|)$ saturation for $Q^2 \gtsim 1\,\GeV^2$
at $x \ltsim 10^{-10}$ which is far below the $x$-regions accessible
at HERA ($x \gtsim 10^{-6}$) and THERA ($x\gtsim 10^{-7}$).

Note that the $S$-matrix unitarity condition together with
relation~(\ref{Eq_xg(x,Q^2,b)-J_gLp(x,b,Q^2)_relation}) requires the
saturation of the impact parameter dependent gluon distribution
$xG(x,Q^2,|\vec{b}_{\!\perp}|)$ but not the saturation of the
integrated gluon distribution $xG(x,Q^2)$. Indeed, approximating
$xG(x,Q^2,|\vec{b}_{\!\perp}|)$ in the saturation regime by a
step-function,
$xG(x,Q^2,|\vec{b}_{\!\perp}|) 
\approx xG^{max}(Q^2)\,\Theta(\,R(x,Q^2)-|\vec{b}_{\!\perp}|\,)$,
where $R(x,Q^2)$ denotes the full width at half maximum of the profile
function, one obtains
\be
        xG(x,Q^2) 
        \;\approx\;
        1.305\,\frac{Q^2\,R^2(x,Q^2)}{\pi \alphaS}\,
        \frac{\pi}{\alphaEM}\,
        J_{\gamma^*_L p}^{max}(Q^2)
        \;\approx\;
        \frac{Q^2\,R^2(x,Q^2)}{\pi\alphaS}
        \ ,
\label{Eq_xg(x,Q^2)_saturation_regime}
\ee
which does not saturate because of the increase of the effective
proton radius $R(x,Q^2)$ with decreasing $x$. Nevertheless, although
$xG(x,Q^2)$ does not saturate, the saturation of
$xG(x,Q^2,|\vec{b}_{\!\perp}|)$ leads to a slow-down in its growth
towards small $x$.

% ___ Conclusion _______________________________________________________________
\section{Conclusion}
\label{Sec_Conclusion}
% ______________________________________________________________________________

We have computed saturation effects in hadronic cross sections with
the loop-loop correlation model (LLCM). The LLCM combines perturbative
and non-perturbative QCD in agreement with lattice investigations,
provides a unified description of $pp$, $\gamma^* p$, and
$\gamma\gamma$ reactions, and respects the $S$-matrix unitarity
condition in impact parameter space. We have calculated impact
parameter profiles of $pp$ collisions in good agreement with the data
for total and differential elastic cross sections. Predictions for
measurements of these cross sections at the LHC were given. While the
effective transverse expansion of the proton continues with increasing
c.m.\ energy, the proton opacity in $pp$ collisions saturates at the
black disc limit for ultra-high energies of $\sqrt{s} \gtsim
10^6\,\GeV$ according to our model. This saturation tames the growth
of the total $pp$ cross section in agreement with the Froissart bound.
We have computed the impact parameter dependent gluon distribution of
the proton $xG(x,Q^2,|\vec{b}_{\!\perp}|)$ from the profile function
for $\gamma^*_L p$ reactions. The corresponding black disc limit is
given by the normalization of the photon wave function and imposes a
unitarity bound on $xG(x,Q^2,|\vec{b}_{\!\perp}|)$.  Accordingly, the
impact parameter dependent gluon distribution
$xG(x,Q^2,|\vec{b}_{\!\perp}|)$ saturates for $Q^2 \gtsim 1\,\GeV^2$
at $x\ltsim 10^{-10}$, which tames the steep rise of the integrated
gluon distribution $xG(x,Q^2)$ towards small~$x$.

% ___ References _______________________________________________________________
%
% ______________________________________________________________________________

%INDEX%%%%%%%%%%%%%%%%%%%%%%%%%%%%%%%%%%%%%%%%%%%%%%%%%%%%%%%%%%%%%%%
% Please check with the editor of your book whether he plans to
% include a "mutual" subject index - if so, please code your entries
% in the standard syntax. For your own purposes you may print your
% "personal" index by using the following commands:
%
%\clearpage
%\addcontentsline{toc}{section}{Index}
%\flushbottom
%\printindex
%%%%%%%%%%%%%%%%%%%%%%%%%%%%%%%%%%%%%%%%%%%%%%%%%%%%%%%%%%%%%%%%%%%%%

\end{document}